\documentclass[pra,twocolumn,showpacs]{revtex4}
\usepackage{amssymb}
\usepackage{amsfonts}
\usepackage{amsmath}
\usepackage{graphicx}
\usepackage{bm}
\usepackage{braket}

\begin{document}

\title{Tunable energy-level inversion in spin-orbit-coupled Bose-Einstein
condensates}
\author{Huan-Bo Luo$^1$, Boris A. Malomed$^{2,3}$, Wu-Ming Liu$^4$, and Lu Li%
$^1$ }
\email{llz@sxu.edu.cn}
\affiliation{$^1$Institute of Theoretical Physics and Department of Physics, State Key
Laboratory of Quantum Optics and Quantum Optics Devices, Collaborative
Innovation Center of Extreme Optics, Shanxi University, Taiyuan 030006, China}
\affiliation{$^2$Department of Physical Electronics, School of Electrical Engineering,
Faculty of Engineering, Tel Aviv University, Tel Aviv 69978, Israel}
\affiliation{$^3$Instituto de Alta Investigaci\'{o}n, Universidad de Tarapac\'{a},
Casilla 7D, Arica, Chile}
\affiliation{$^4$Beijing National Laboratory for Condensed Matter Physics, Institute of
Physics, Chinese Academy of Sciences, Beijing 100190, China}

\begin{abstract}
A method to realize controllable inversion of energy levels in a
one-dimensional spin-orbit (SO)-coupled two-component Bose-Einstein
condensate under the action of a gradient magnetic field and
harmonic-oscillator (HO) trapping potential is proposed. The linear version
of the system is solved exactly. By adjusting the SO coupling strength and
magnetic-field gradient, the energy-level inversion makes it possible to
transform any excited state into the ground state. The full nonlinear system
is solved numerically, and it is found that the results are consistent with
the linear prediction in the case of the repulsive inter-component
interaction. On the other hand, the inter-component attraction gives rise to
states of superposition and edge types. Similar results are also reported
for the system with the HO trap replaced by the box potential. These results
suggest a possibility to realize any excited state and observe it in the
experiment.
\end{abstract}

\maketitle

\section{introduction}

Atomic Bose-Einstein condensates (BECs) are easily-tunable quantum
macroscopic systems, which offer an ideal experimental platform for
simulating various effects known in condensed-matter physics~\cite%
{RepProgPhys.75.082401,Lewenstein}. A well-known example is the spin-orbit
(SO) coupling in semiconductors, which plays a fundamental role in the
realization of spin Hall effects~\cite{RevModPhys.82.1959}, topological
insulators~\cite{RevModPhys.82.3045}, spintronic devices~\cite%
{RevModPhys.76.323}, \textit{etc}. Since the emulation of the SO coupling in
effectively one-dimensional (1D)~\cite{nature09887,Juzeliunas} and
two-dimensional~(2D)~\cite{Science.354.83} BEC was implemented in the
experiment, many remarkable effects in SO-coupled BECs with intrinsic
nonlinearity have been predicted by numerically solving the respective
Gross-Pitaevskii equations~\cite{BEC-SOC GP eqns}, such as vortices~\cite%
{Kawakami,Drummond,PhysRevLett.109.015301,Sakaguchi}, skyrmions~\cite%
{PhysRevLett.109.015301} and various species of solitons~\cite%
{PhysRevLett.110.264101,1D sol 2,1D sol 3,1D sol
4,PhysRevE.89.032920,PhysRevE.94.032202,Cardoso,Lobanov,2D SOC gap sol
Raymond,SOC 2D gap sol Hidetsugu,low-dim SOC,Han Pu 3D}, see also reviews of
the experimental and theoretical findings in Refs.~\cite%
{Spielman,Galitski,Ohberg,Zhai,SOC-sol-review}. However, energy-level
inversion in BEC, which, as we demonstrate in this work, can be induced by
the combined effect of a gradient magnetic field and SO coupling, has not
been found previously.

The energy-level quantization is a commonly known feature of spatially
confined quantum-mechanical systems, such as atomic BECs trapped in an
external potential. Due to the lack of a mechanism for rearrangement of
energy levels, most studies have been performed at the lowest-energy (i.e.,
ground-state, GS)~level~\cite%
{Kawakami,Drummond,PhysRevLett.109.015301,Sakaguchi}. To relax this
restriction, a tunable energy-level inversion mechanism, which alters the
spectrum of eigen-energies but does not essentially affect the corresponding
eigenfunctions, may be very relevant. Ideally, it should be possible to
transform any excited state into the GS, which will open the way to realize
excited states in the experiment. To this end, states with higher quantum
numbers, similar to Rydberg ones, can be addressed. Besides that, the
transition mechanism between different energy levels is of interest in its
own right.

In this paper, we propose a method to realize a tunable energy-level
inversion in the SO-coupled BEC under the action of a gradient magnetic
field and harmonic-oscillator (HO) trapping potential. By introducing a
shifted quantum-number density operator, the linear version of the system is
solved exactly. In this case, the combined effect of the SO coupling and
gradient magnetic field can reduce the total energy, so that the higher the
energy level is the more it drops. Thus, by adjusting the SO coupling
strength and magnetic-field gradient, one can realize the energy-level
inversion, making it possible, indeed, to convert any excited state into the
GS. In addition to the exact solution of the linear system, its nonlinear
counterpart, including both repulsive and attractive inter-component
interactions, is solved numerically.

The bulk of the paper is structured as follows. In Sec. II, the theoretical
model is introduced. In Sec. III, the linear solution is constructed in
terms of a pair of Hermite-Gaussian functions. In Secs. IV and V, numerical
solutions of the nonlinear system with repulsive and attractive
inter-component interactions are addressed. In Sec. VI, the numerical
solution is presented for the system with the HO trapping potential replaced
by a box-shaped one. Findings produced by this work are summarized in Sec.
VII.

\section{The model and its reductions}

We consider the SO-coupled effectively 1D binary BEC under the action of the
normalized HO potential, $V(x)=x^{2}/2$, and dc magnetic field, $\mathbf{B}%
(x)=\{-(\alpha /\sqrt{2})x,0,\Omega \}$, with constant gradient $-\alpha /%
\sqrt{2}$ along the $x$ direction and a uniform magnetic field $\Omega $ in
the $z$ direction. The SO coupling is chosen as~\cite{nature09887} $V_{\text{%
so}}=i(\beta /\sqrt{2})\sigma _{y}\partial _{x}$, where $\sigma =(\sigma
_{x},\sigma _{y},\sigma _{z})$ is the vector of the Pauli matrices, and $%
\beta $ is a parameter which determines the SO coupling strength. The spinor
wave function, $\Psi =(\Psi _{1},\Psi _{2})^{T}$, obeys the respective
system of 1D Gross-Pitaevskii equations, whose scaled form is
\begin{equation}
\begin{split}
i\partial _{t}\Psi _{1}=& \frac{1}{2}\left( -\partial _{x}^{2}+x^{2}\right)
\Psi _{1}-\frac{1}{\sqrt{2}}\left( \alpha x-\beta \partial _{x}\right) \Psi
_{2} \\
& +\Omega \Psi _{1}+\left( g|\Psi _{1}|^{2}+\gamma |\Psi _{2}|^{2}\right)
\Psi _{1}, \\
i\partial _{t}\Psi _{2}=& \frac{1}{2}\left( -\partial _{x}^{2}+x^{2}\right)
\Psi _{2}-\frac{1}{\sqrt{2}}\left( \alpha x+\beta \partial _{x}\right) \Psi
_{1} \\
& -\Omega \Psi _{2}+\left( g|\Psi _{2}|^{2}+\gamma |\Psi _{1}|^{2}\right)
\Psi _{2},
\end{split}
\label{main}
\end{equation}%
where $g$ and $\gamma $ are coefficients of the intra- and inter-component
interactions, respectively. We set below
\begin{equation}
\Omega =\beta \Delta -1/2,  \label{Omega}
\end{equation}
with $\Delta $ being the quantum-number shift. Using the remaining scaling
invariance of Eq.~\eqref{main}, we fix $g=1$, which assumes, as usual, the
repulsive sign of the self-interaction of each component (while $\gamma $
may be negative, accounting for attraction between the components, which can
be induced by means of the Feshbach resonance~\cite{Feshbach}). Here we
assume nearly equal parameters of the magnetic fields, $\alpha $ and $\beta $
[in Eq. (\ref{Omega})], i.e.,
\begin{equation}
\alpha =\beta +\delta \beta ,  \label{ab}
\end{equation}
where $\delta \beta $ is a small constant. Thus, only $\beta $, $\Delta $
and $\gamma $ are kept as free parameters of the system, while effects of $%
\delta \beta $ are negligible.

Equations~\eqref{main} are written in the scaled form. In physical units,
assuming that the binary condensate is a mixture of two different atomic
states of $^{87}$Rb \cite{nature09887}, relevant values of the trapping
frequency are $\omega=10$Hz. The number of atoms in the condensates is $1000$. 
This number of atoms is sufficient to
observe the predicted patterns in the experiment in full detail. The 
characteristic length, time and energy are defined by $l =\sqrt{\hbar /m\omega}=8.55\mu$m, $\tau=1/\omega=100$ms and $\epsilon= \hbar \omega=1.05\times10^{-33}$J, 
where $m = 1.44\times10^{-25}$kg is the mass of $^{87}$Rb.

Stationary solutions of Eq.~\eqref{main} with chemical potential $\mu $ are
sought for in the usual form, $\Psi =\psi \exp (-i\mu t)$ and $\psi =\{\psi
_{1}(x),\psi _{2}(x)\}^{T}$. Note that Eq.~\eqref{main} is compatible with
substitution
\begin{equation}
\psi _{1}(x)\rightarrow \psi _{1}(-x),\ \ \psi _{2}(x)\rightarrow -\psi
_{2}(-x),  \label{substitution}
\end{equation}%
which means that the system admits self-conjugate solutions, subject to the
symmetry constraint%
\begin{equation}
|\psi _{1}(x)|^{2}=|\psi _{1}(-x)|^{2},\ \ |\psi _{2}(x)|^{2}=|\psi
_{2}(-x)|^{2},  \label{symmetriy}
\end{equation}%
or a pair of degenerate solutions related by transformation (\ref%
{substitution}) if the self-conjugation (symmetry) is broken by the
self-attractive nonlinearity.

\section{Exact solutions of the linearized system}

\begin{figure}[tbp]
\centering
\includegraphics[width=3.2in]{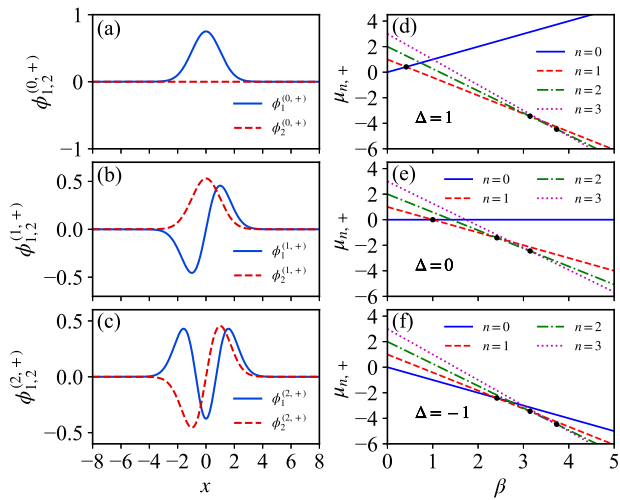}
\caption{(Color online) (a-c): Profiles of the normalized wave functions $\protect\phi %
_{1,2}^{(n,+)}$ with quantum number (a) $n=0$, (b) $n=1$ and (c) $n=2$ with $%
\Delta =0$. (d-f): Linear chemical potential $\protect\mu _{n,+}(\protect%
\beta ,\Delta )$ with (d) $\Delta =1$, (e) $\Delta =0$ and (f) $\Delta =-1$,
plotted pursuant to Eq.~\eqref{mu}. The dots are values of $\protect\beta %
_{n}$ given by Eq.~\eqref{beta}.}
\label{figure1}
\end{figure}

First, we note that the~stationary linear version of Eq. \eqref{main} with $%
\delta \beta \!=\!0$ [see Eq. (\ref{ab})], i.e., $\hat{H}\psi \!=\!\mu \psi $
with the linear Hamiltonian,
\begin{equation}
\hat{H}=\frac{1}{2}\left( \!-\partial _{x}^{2}\!+\!x^{2}\!-\!\sigma
_{z}\!\right) \!-\!\beta \!\left[ \!\frac{1}{\sqrt{2}}\left( x\sigma
_{x}-i\partial _{x}\sigma _{y}\!\right) \!-\!\Delta \sigma _{z}\!\right] ,
\label{Hamiltonian}
\end{equation}%
admits an exact solution. Indeed, in terms of the shifted quantum-number
density operator,
\begin{equation}
\hat{P}\!=\!(x\sigma _{x}\!-\!i\partial _{x}\sigma _{y})\!/\!\sqrt{2}%
\!-\!\Delta \sigma _{z},  \label{P}
\end{equation}
the Hamiltonian can be written as
\begin{equation}
\hat{H}\!=\!\hat{P}^{2}\!-\beta \hat{P}\!-\Delta ^{2}.  \label{H}
\end{equation}
Then, the solutions of the auxiliary eigenvalue equation,
\begin{equation}
\hat{P}\Phi _{n,\pm }\!=\!\rho _{n,\pm }\Phi _{n,\pm },  \label{P2}
\end{equation}
with real eigenvalue $\rho _{n,\!\pm }$ and eigenstate $\Phi _{n,\!\pm }$$%
=\{\phi _{1}^{(n,\pm )}(x),\phi _{2}^{(n,\pm )}(x)\}^{T}$, can be found in
the form similar to HO\ eigenstates \cite{LL}:
\begin{equation}
\begin{split}
\phi _{1}^{(n,\pm )}& =\frac{1}{A_{n,\pm}}H_{n}(x)\exp \left( -\frac{x^{2}}{2}\right) , \\
\phi _{2}^{(n,\pm )}& =\frac{\sqrt{2}}{A_{n,\pm}}(\rho _{n,\pm }+\Delta )H_{{n}-1}(x)\exp
\left( -\frac{x^{2}}{2}\right),
\end{split}
\label{psi}
\end{equation}%
where the standard Hermite 
polynomials are
\begin{equation}
H_{n}(x)\equiv (-1)^{n}\exp (x^{2})\frac{d^{n}}{dx^{n}}\exp (-x^{2}),
\label{Hermite}
\end{equation}
with the quantum number $n=0,1,2,\cdots $. For $n=0$, 
we set $H_{-1}\equiv 0$ in Eq.~\eqref{psi}. The respective eigenvalues,
produced by Eqs. (\ref{P2}) and (\ref{P}), being
\begin{equation}
\rho _{n,\pm }=\left\{
\begin{array}{ll}
-\Delta , & n=0, \\
\pm \sqrt{n+\Delta ^{2}}, & n=1,2,3\cdots .%
\end{array}%
\right.  \label{rho}
\end{equation}%
The normalization coefficients $A_{n,\pm}$ are defined by
\begin{equation}
    A^2_{n,\pm}=\left\{
    \begin{array}{ll}
    \sqrt{\pi} , & n=0, \\
    2^{n+1}\sqrt{\pi}(\rho_{n,\pm}^2+\Delta \rho_{n,\pm})(n-1)!, &n=1,2\cdots .%
    \end{array}%
    \right.  \label{A}
\end{equation}%
Then, eigenstates~%
\eqref{psi} are built as pairs of the Hermite-Gaussian functions of orders $%
n $ and $n-1$ in the two components, whose typical profiles are shown in
Figs.~\ref{figure1}(a)-(c). Each node (zero) of $\phi _{1}^{n,+}$
corresponds to the peak or valley of $\phi _{2}^{n,+}$, which implies that
the solutions feature the structure with spatially separated components.

Because operator $\hat{P}$ commutes with $\hat{H}$, the eigenstate given by
Eq.~\eqref{psi} is also an eigenstate of $\hat{H}$, with the respective
chemical potential
\begin{equation}
\begin{split}
& \mu _{n,\pm }(\beta ,\Delta )=\rho _{n,\pm }^{2}-\beta \rho _{n,\pm
}-\Delta ^{2} \\
& =\left\{
\begin{array}{ll}
\beta \Delta , & n=0, \\
n\mp \beta \sqrt{n+\Delta ^{2}}, & n=1,2,3,\cdots ,%
\end{array}%
\right.
\end{split}
\label{mu}
\end{equation}%
which is a function of $\beta $ and $\Delta $ [recall $\beta $ is defined as
per Eq. (\ref{Omega})]. Results below are presented only for eigenvalues, $%
\mu _{n,+}$, as they are lower than $\mu _{n,-}$. One can see that parameter
$\beta $ alters the energy spectra but does not affect the corresponding
eigenfunctions. Thus, it is relevant to discuss the effect of $\beta $ on
the energy, aiming to find out where the energy-level inversion occurs.

Figures~\ref{figure1}(d-f) present the dependence of chemical potential $\mu
_{n,+}$ on $\beta $ at $\Delta =0,\pm 1$. It is seen that $\mu _{n,+}=n$ at $%
\beta =0$, which means that the GS corresponds to $n=0$ in this case. The
situation is different for $\beta \neq 0$. With the increase of $\beta $,
eigenvalues $\mu _{n,+}(\beta )$ and $\mu _{n+1,+}(\beta )$ collide,
switching their ordering from $\mu _{n,+}(\beta )<\mu _{n+1,+}(\beta )$ to $%
\mu _{n+1,+}(\beta )<\mu _{n,+}(\beta )$, at critical values of the SO
coupling strength
\begin{equation}
\beta _{n}(\Delta )\!=\!\left\{
\begin{array}{ll}
\sqrt{1+\Delta ^{2}}-\Delta , & n=0, \\
\sqrt{n\!+\!1\!+\Delta ^{2}}\!+\sqrt{n\!+\!\Delta ^{2}}, & n=1,2,\cdots .%
\end{array}%
\right.   \label{beta}
\end{equation}%
In particular, values $\beta _{0}=1$, $\beta _{1}\approx 2.414$ and $\beta
_{2}\approx 3.146$, given by Eq.~\eqref{beta} with $\Delta =0$, are marked
by dots in Fig.~\ref{figure1}(e). Thus, the state with $n=0$ is the GS at $%
0<\beta <\beta _{0}$, while the state with $n\geq 1$ becomes the GS at $%
\beta _{n-1}<\beta <\beta _{n}$. Accordingly, the energy-level inversion
which occurs at $\beta =\beta _{n}(\Delta )$ may be considered as the GS
phase transition.

\section{Numerical results under inter-component repulsive interaction}

Next, we consider the complete form of Eq.~\eqref{main} including the
nonlinear interactions, repulsive or attractive. In this case, stationary
states can be found in a numerical form by means of the imaginary-time
propagation method. For this purpose, we fix the total norm as $%
N=\left\langle \psi |\psi \right\rangle =1$.

\begin{figure}[tbp]
\centering
\includegraphics[width=3.2in]{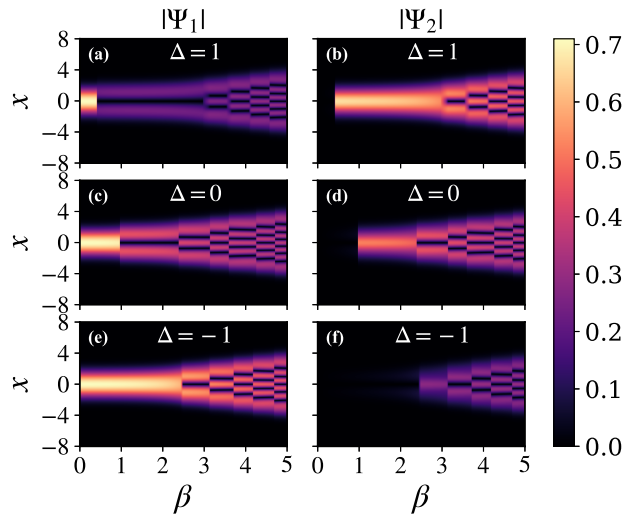}
\caption{(Color online) Distributions of the absolute values of the wave
functions in the two components of the GS for (a,b) $\Delta =1$, (c,d) $%
\Delta =0$, and (e,f) $\Delta =-1$. Here the nonlinearity coefficients in
Eq.~\eqref{main} are $\protect\gamma =3$ and $g=1$.}
\label{figure2}
\end{figure}

\begin{figure}[tbp]
\centering
\includegraphics[width=3.2in]{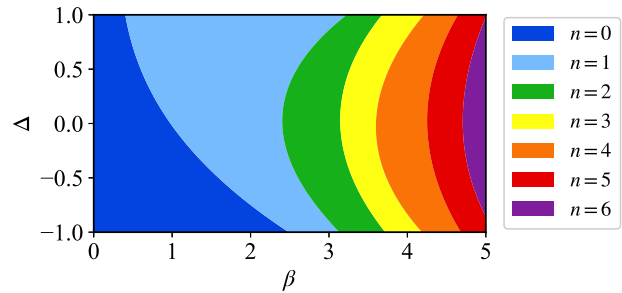}
\caption{(Color online) Map of values of quantum number $n$ corresponding to
the GSs of the nonlinear system with the inter-component repulsion, in the $%
(\Delta ,\protect\beta )$ plane. Here the nonlinearity coefficients in Eq.~
\eqref{main} are $\protect\gamma =3$ and $g=1$.}
\label{figure3}
\end{figure}

We start with the case of the inter-component repulsion, i.e., $\gamma >0$
in Eq.~\eqref{main}. In this case, the wave function tends to feature
spatial separation between the two components, similar to wave functions~%
\eqref{psi} of the linear system. Figure~\ref{figure2} shows the numerical
results for $\gamma =3$, $\beta $ ranging from $0$ to $5$, and $\Delta
=-1,0,+1$. With the increase of $\beta $, the GS is carried over from one
corresponding to $n$ to the adjacent state, with quantum number $n+1$, also
similar to the situation in the linear system. As $n\geq 3$, the nonlinear
GS develops a pattern in the form of a spatially confined lattice, in both $%
\psi _{1}$ and $\psi _{2}$ components. It is relevant to mention the
asymptotic expression for the Hermite-Gaussian functions with $n\rightarrow
\infty $:
\begin{equation}
H_{n}(\!x\!)\exp \!\left( \!\!-\frac{x^{2}}{2}\!\right) =C\cos \!\left( \!x%
\sqrt{2n}\!-\frac{\pi n}{2}\!\right) \!\!\left( \!1\!-\!\frac{x^{2}}{2n\!}%
\right) ,  \label{asymptotic}
\end{equation}%
which is valid at $|x|<\sqrt{2n}$, and $C$ is a real constant. With the help
of the asymptotic expression~\eqref{asymptotic}, the period and size of the
lattice can be approximated by $T_{n}=\pi \sqrt{2/n}$ and $L_{n}=\sqrt{2n}$,
for both components.

Once again similar to the linear system, parameter $\Delta $ exerts two
effects on the GSs. One is to adjust the ratio of norms in the two
components, as shown in Fig.~\ref{figure2}. The other effect of $\Delta $ is
to shift the phase-transition point, as seen in the phase diagram displayed
in Fig.~\ref{figure3}. With the increase of $\Delta $, value $\beta _{0}$ of
the SO coupling strength at the first phase-transition point tends to vanish
at $\Delta \rightarrow +\infty $. As for the states with $n\geq 1$, the
phase-transition points are almost symmetric with respect to $\Delta =0$.
These results also resemble the above findings for the linear system
presented in Eq.~\eqref{beta}.

\begin{figure}[tbp]
\centering
\includegraphics[width=3.2in]{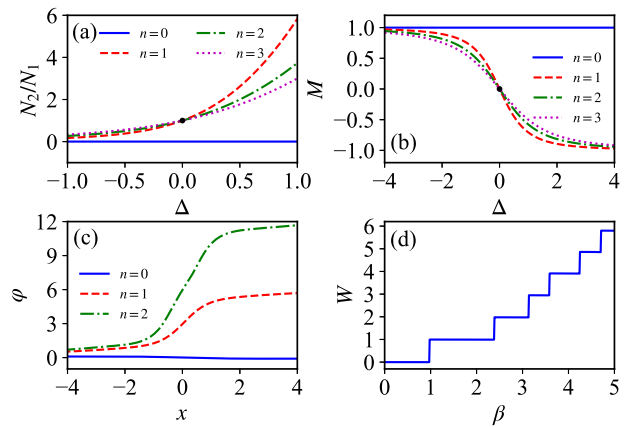}
\caption{(Color online) (a) Ratio $N_{2}/N_{1}$ as a function of $\Delta $
and (b) the magnetization curves for $n=0,1,2$ and $3$. (c) The spatial
profile of phase $\protect\varphi $ for $n=0,1,2$ and $3$ at $\Delta =0$.
(d) Winding number $W$, defined by Eq.~\eqref{Winding}, for $\protect\beta $
ranging from $0$ to $5$ at $\Delta =0$.}
\label{figure4}
\end{figure}

The particle numbers (norms) of each component can be expressed as $%
N_{1}=\int_{-\infty }^{+\infty }|\psi _{1}|^{2}dx$ and $N_{2}=\int_{-\infty
}^{+\infty }|\psi _{1}|^{2}dx$. Further, the ratio of the norms of the two
components, $N_{1}/N_{2}$, is plotted in Fig.~\ref{figure4}(a) as a function
of $\Delta $ with quantum number $n$, and can be approximated by
\begin{equation}
\frac{N_{2}}{N_{1}}=\left\{
\begin{array}{ll}
0, & n=0, \\
(\sqrt{n+\Delta ^{2}}+\Delta )^{2}/n, & n=1,2,3\cdots ,%
\end{array}%
\right.   \label{ratio}
\end{equation}%
which is the ratio for the solutions~\eqref{psi} of the linearized
equations. It is relevant to mention that the two-component Bose gas can be
considered as a (pseudo-)spin system. The spin vector, $\mathbf{S}=\psi
^{\dagger }\bm{\sigma}\psi /\psi ^{\dagger }\psi $, can be used to represent
the respective pseudo-magnetic ordering. The corresponding magnetization $M$
can be defined by the average $z$-component of the spin,
\begin{equation}
M\equiv \bar{S}_{z}=\frac{\left\langle \psi \right\vert \sigma
_{z}\left\vert \psi \right\rangle }{N_{1}+N_{2}}=\frac{1-N_{2}/N_{1}}{%
1+N_{2}/N_{1}}.  \label{M1}
\end{equation}%
Substituting expression~\eqref{ratio} in Eq.~\eqref{M1} yields
\begin{equation}
M=\left\{
\begin{array}{ll}
1, & n=0, \\
-\Delta /\sqrt{n+\Delta ^{2}}, & n=1,2,3\cdots .%
\end{array}%
\right.   \label{M2}
\end{equation}%
The magnetization curves for $n=0,1,2,3$ are plotted in Fig.~\ref{figure4}%
(b).

Here we only discuss the properties of the magnetization curves when $n\neq 0
$. Without the effective magnetic field applied to the BEC, i.e., at $\Delta
=0$, atoms are evenly distributed in the $\psi _{1}$ and $\psi _{2}$
components, i.e. $N_{2}/N_{1}=1$, hence the magnetization vanishes. With the
increase of the effective magnetic field $\Delta $, the atomic population is
transferred from $\psi _{1}$ to $\psi _{2}$, which yields a lower
magnetization. Eventually, at the critical value of the effective field, $%
\Delta _{n}=2\sqrt{n}$, nearly all the atoms are transferred to $\psi _{2}$.
The magnetization remains nearly constant, i.e. $-1<M<-0.9$, which means
saturation of the magnetization. This result indicates that the critical
value of the magnetization, corresponding to the saturation, grows with the
increase of quantum number $n$.

Considering the eigenstates of the linear system given by Eq.~\eqref{psi},
the coordinate axis of $x$ can be mapped into circle $S^{1}$, with elements $%
\mathbf{S}(x=-\infty )=\mathbf{S}(x=+\infty )=(0,0,1)$. The manifolds of the
spin vector is also $S^{1}$, as $S_{y}=0$ and $S_{x}^{2}+S_{z}^{2}=1$. Thus,
the distribution of the spin vector of the eigenstates can be classified by
the fundamental homotopy group, $\pi _{1}(S^{1})=\mathbb{Z}$, being
characterized by the winding number as follows:
\begin{equation}
W=\frac{1}{2\pi }\int_{-\infty }^{+\infty }d\varphi (x)=\frac{\varphi
(+\infty )-\varphi (-\infty )}{2\pi },  \label{Winding}
\end{equation}%
where $S_{x}+iS_{z}\equiv i\exp (i\varphi )$. Note that if we define $%
\varphi (x=-\infty )=0$, then $\varphi (x=+\infty )=2n\pi $, hence the
winding number of the eigenstate is $W=n$. The results for the phase pattern
$\varphi (x)$ and winding numbers $W$ for the linear eigenmodes are shown in
Figs.~\ref{figure4}(c,d). This result implies that the GS phase transition
points, produced by Eq.~\eqref{beta}, also yield values of the winding
number at the phase-transition points. The winding numbers correspond to the
number of zeros of the $\psi _{1}$, or the number of peaks of $|\psi _{2}|$.
One reason for the emergence of the GS phase transition is that the
solutions with different winding numbers cannot be transformed into each
other by continuous deformations. Thus, a general conclusion is that the
eigenstates of the nonlinear system with the repulsive inter-component
interaction are quite similar to their counterparts produced by the linear
system in the previous section.

\section{Numerical results under inter-component attractive interaction}

\begin{figure}[tbp]
\centering
\includegraphics[width=3.2in]{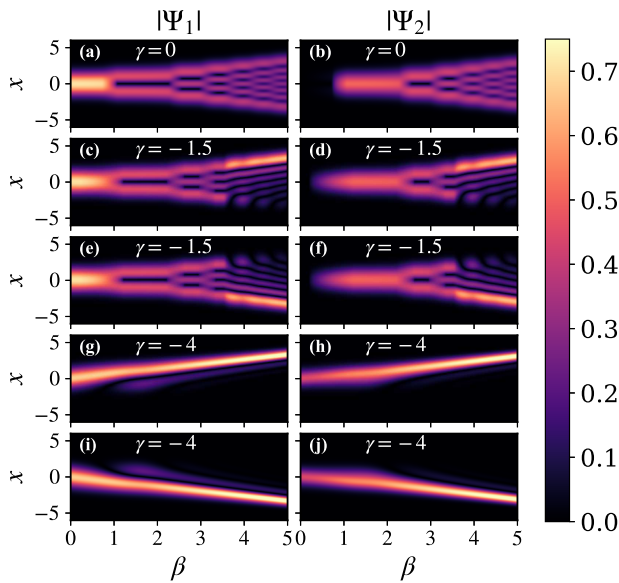}
\caption{(Color online) Distributions of the absolute value of the wave
functions in the two components of the GS for (a,b) $\protect\gamma =0$,
(c-f) $\protect\gamma =-1.5$, and (g-j) $\protect\gamma =-4$. Here $\Delta =0
$, and the self-repulsion coefficient in Eq.~\eqref{main} is $g=1$.}
\label{figure5}
\end{figure}

Next, we consider the nonlinear system with the inter-component attraction,
i.e., $\gamma \leq 0$ in Eq.~\eqref{main}. Figure~\ref{figure5} displays the
variation of the density distribution in the respective numerically found
GS, driven by the increase of $\beta $ at $\Delta =0$. The figure exhibits
completely different patterns with the increase of $|\gamma |$. For $\gamma
=0$, a superposition state appears near the linear phase-transition points $%
\beta _{n}$ in Figs.~\ref{figure5}(a,b), cf. similar patterns displayed by
Fig. \ref{figure2} for the nonlinear system with $\gamma >0$. For $\gamma
=-1.5$, the GS takes the form of an edge state at $\beta >3.5$. Note that
this state breaks the spatial symmetry defined by Eq.~\eqref{symmetriy},
which implies the existence of a pair of degenerate states, i.e., top and
bottom edge state, see Figs.~\ref{figure5}(c,d) and (e,f). For $\gamma =-4$,
the GS takes the shape of the edge state for all values of $\beta >0$, see
Figs.~\ref{figure5}(g,h) and (i,j).

To distinguish the eigenstates of three types, i.e., the simple one, the
superposition pattern, and the edge state, we focus on the case of $\gamma
=-1.5$, when the top edge state is observed in Figs.~\ref{figure5}(c,d). All
the states can be expressed as a superposition of all linear eigenstates~%
\eqref{psi}:
\begin{equation}
\psi =\sum_{\pm }\sum_{n=0}^{\infty }c_{n,\pm }\Phi
_{n,\pm },c_{n,\pm }=\int_{-\infty }^{+\infty }\Phi _{n,\pm }^{\dagger
}\psi dx,  \label{superposition}
\end{equation}%
where the coefficients $c_{n,\pm }$  satisfy $\sum_{\pm
}\sum_{n=0}^{\infty }|c_{n,\pm }|^{2}=1$. Further, we can define $c_{n,\pm
}=|c_{n,\pm }|\exp (i\theta _{n,\pm })$, where $\theta _{n,\pm }$ are
phases, while $|c_{n,\pm }|^{2}$ accounts for the weight of each eigenstate.

\begin{figure}[tbp]
\centering
\includegraphics[width=3.2in]{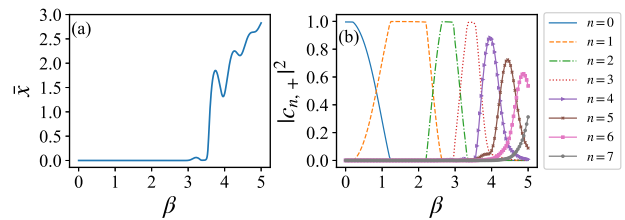}
\caption{(Color online) Dependence of (a) the average displacement $\bar{x}$
of the GS, and (b) weights $|c_{n,+}|^{2}$ of each linear eigenstate in the
expansion~\eqref{superposition} of this GS on $\protect\beta $. Coefficients
of Eq.~\eqref{main} are $\protect\gamma =-1.5$, $\Delta =0$ and $g=1$.}
\label{figure6}
\end{figure}

Figure~\ref{figure6}(b) shows the dependence of weight $|c_{n,\pm }|^{2}$ on
$\beta $ for $\gamma =-1.5$. It is clearly seen that, with the increase of $%
\beta $, the GS is, initially, similar to the linear eigenstate with $n=0$,
having $|c_{0,+}|^{2}=1$. As $0.3<\beta <1.2$, the GS develops the shape of
the superposition of two linear eigenstates with $n=0$ and $n=1$. Further
increasing $\beta $, the GS becomes similar to the linear eigenstate with $%
n=1$, having $|c_{1,+}|^{2}=1$. Thus, in the region of $0.3<\beta <1.2$, the
GS features a transition from the nearly linear eigenstate with $n=0$ to the
one with $n=1$. In general, the GS phase transition occurs close to the
critical value $\beta _{n}$ [see Eq.~\eqref{beta}], near which the wave
function is close to the superposition of two linear eigenstates with
quantum numbers $n$ and $n+1$, the phase difference between which is $%
|\theta _{n+1}-\theta _{n}|=\pi /2$. Thus, the spatial density of the wave
function may be approximated by weighted sum, i.e., $|\psi
_{j}|^{2}\!=\!|c_{n,+}\!|^{2}|\phi
_{j}^{(n,+)}\!|^{2}\!+\!|\!c_{n+1,+}\!|^{2}|\phi
_{j}^{(n+1,+)}\!|^{2},j\!=\!1,2$. Therefore the GS of the superposition type
may still feature the symmetry defined by Eq.~\eqref{symmetriy}. At $\beta
>3.5$, the GS is the superposition of more than two linear eigenstates~%
\eqref{psi} with equal phases of all the constituents, which takes the form
of edge states, breaking the symmetry defined by Eq.~\eqref{symmetriy}. The
symmetry breaking is characterized by the dependence of $\bar{x}$ on $\beta $%
, as shown in Fig.~\ref{figure6}(a), where $\bar{x}=\int_{-\infty }^{+\infty
}\psi ^{\dagger }x\psi dx$ is the average displacement.

For $\gamma \!=\!-4$, the inter-component attraction is the dominant factor
in the system, resulting in $\psi _{1}\!\approx \!\pm \psi _{2}$, while the
SO coupling and Zeeman splitting may be omitted, as their energies, $E_{%
\text{soc}}\!=\!(\beta \!/\!\sqrt{2})\int_{-\infty }^{+\infty }(\psi
_{1}\partial _{x}\psi _{2}-\psi _{2}\partial _{x}\psi _{1})dx$ and $%
E_{Z}=\Omega \int_{-\infty }^{+\infty }(|\psi _{1}|^{2}-|\psi _{2}|^{2})dx$,
become vanishingly small. Substituting these approximations into Eq.~%
\eqref{main}, one can reduce it to the single-component equation for the
stationary wave function with chemical potential $\mu $,
\begin{equation}
\mu \psi _{1}=\frac{1}{2}(-\partial _{x}^{2}+x^{2}\mp \sqrt{2}\beta x)\psi
_{1}+(1+\gamma )\psi _{1}^{3},  \label{mu1}
\end{equation}
which features an effective potential $\widetilde{V}(x)=x^{2}/2\mp \beta x/%
\sqrt{2}$. The corresponding condensate is localized around minima of the
effective potential, $x_{0}=\pm \beta /\sqrt{2}$, which determine edges of
the above-mentioned spatially confined lattice. With the increase of $\beta $%
, the peaks of density become sharp, as seen in Figs.~\ref{figure5}(g,h) and
(i,j).

\begin{figure}[tbp]
    \centering
    \includegraphics[width=3.2in]{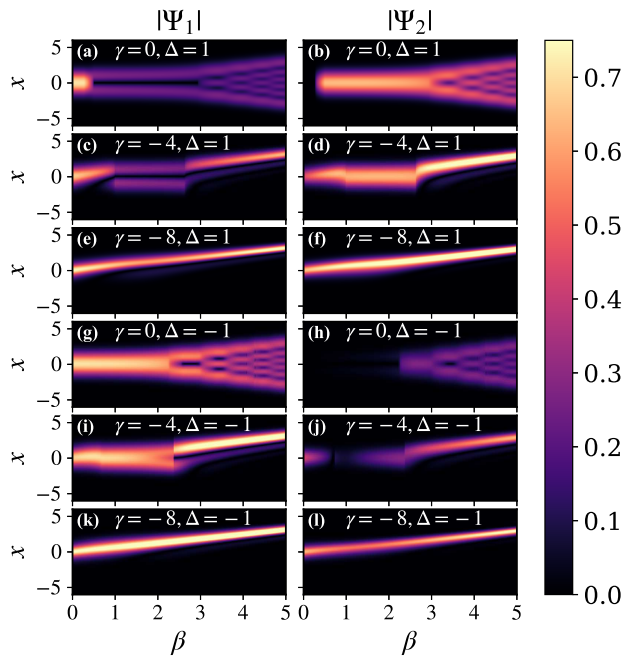}
    \caption{(Color online) Distributions of the absolute value of the wave
    functions in the two components of the GS for (a-f) $\Delta =1$ and (g-l) $%
    \Delta =-1$. Here the self-repulsion coefficient in Eq.~\eqref{main} is $g=1$%
    .}
    \label{figure7}
\end{figure}

The results discussed above are all based on $\Delta =0$. Here we will
discuss how parameter $\Delta $ [defined in Eq.~(\ref{Omega})] affects the
GS in the case of the inter-component attractive interaction. The
corresponding density distributions of the wave functions are shown in Fig.~%
\ref{figure7}. The results can be explained by the combined effect of $%
\gamma $ and $\Delta $. With the increase of $\Delta $, the phase-transition
points are shifted and the particle numbers (norms) of each component are
adjusted. For $\gamma =0$, the superposition state appears near the phase
transition points, see Figs.~\ref{figure7} (a,b) and (g,h). According to Eq.~%
\eqref{beta}, the states with $n=1$ and $n=0$ have a wider range as the GSs
for $\Delta =1$ and $\Delta =-1$, respectively. By comparing Figs.~\ref%
{figure6}(g,h) with Figs.~\ref{figure7}(c,d) and (i,j), one finds that the
states with $n=1,\Delta =1$ and $n=0,\Delta =-1$ at $\gamma =-4$ do not yet
become edge modes. Under the action of stronger inter-component attraction,
at $\gamma =-8$, the GS takes the shape of the edge state for all values of $%
\beta >0$, similar to the case of $\Delta =0$. Note that bottom edge states
(not shown here) can be obtained by substitution~\eqref{substitution},
featuring the same properties as the top edge states.

The above-mentioned results indicate that attractive inter-component
interaction leads to symmetry breaking and the appearance of edge
states, which is completely different from the case of repulsion interaction.
The repulsive interaction tends to form spatially separated states, 
while the linear states~\eqref{psi} happen to be spatially separated. 
Note that the peaks and valleys of $\psi_1$ correspond to nodes of 
$\psi_2$ for linear eigenstates. Therefore, the solutions under repulsive 
interaction are similar to the linear eigenstates. Note that the linear
eigenstates are symmetric. On the contrary, the attractive interaction 
tends to form spatial mixed states, i.e. $\psi_1=\pm\psi_2$, which will 
lead to the competition between the linear part and the nonlinear part. 
With the enhancement of attractive interaction, the edge states (mixed 
states) have lower energy than the linear eigenstates. Thus the edge states
become the ground states and the symmetry is broken. At the same time, 
parameter $\Delta$ has the effect of adjusting the proportion of two 
components, i.e. $\int |\psi_1|^2dx<\int |\psi_2|^2dx$ for $\Delta>0$, 
and vice versa, which is contradictory to the formation of mixed states. 
Therefore, in the case of $\Delta\neq0$, the edge states will not 
become the ground states until there is a stronger attractive interaction.

\begin{figure}[tbp]
    \centering
    \includegraphics[width=3.2in]{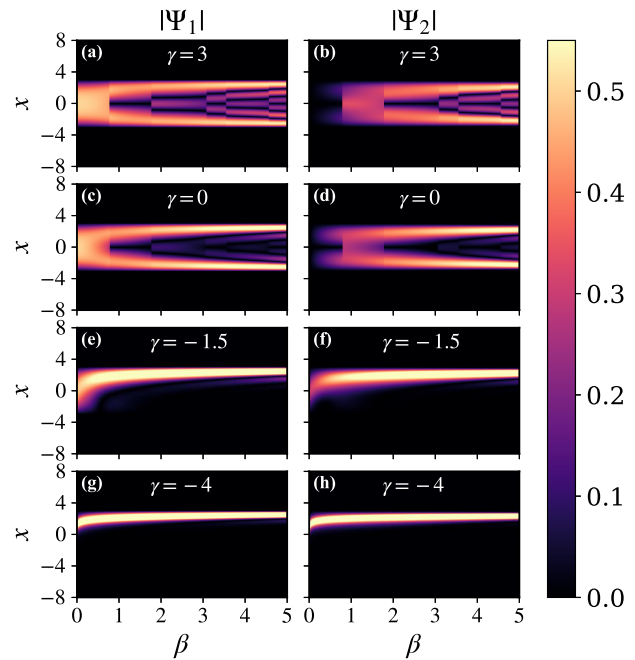}
    \caption{(Color online) Distributions of the absolute value of the wave
    functions in the two components of the GS for (a,b) $\protect\gamma =3$,
    (c,d) $\protect\gamma =0$, (e,f) $\protect\gamma =-1.5$ and (g,h) $\protect%
    \gamma =-4$, for the system with the box trapping potential defined as per
    Eq.~\eqref{box}. The parameters are $\Delta =0$ [see Eq. (\protect\ref{Omega}%
    )] and $g=1$ [the self-repulsion coefficient in Eq.~\eqref{main}].}
    \label{figure8}
    \end{figure}

\section{Numerical results for a box potential}

The above-mentioned results are based on the HO potential in Eq. (\ref{main}%
), which makes it possible to find the exact solution for the linear system.
To investigate the sensitivity of the results to shape of the trap, we here
consider the box potential, defined as
\begin{equation}
V=\left\{
\begin{array}{ll}
-1000, & |x|\leq 3, \\
0, & \mathrm{at}~|x|>3.%
\end{array}%
\right.   \label{box}
\end{equation}%
As well as its HO counterpart, the trap in the form of a deep potential box
was used in experiments with BEC \cite{Hadzi}. In this case, the results,
produced by numerical solution of Eq.~\eqref{main}, are shown in Fig.~\ref%
{figure8}.

The wave function in Figs.~\ref{figure8}(a,b) corresponds to the system with
the inter-component repulsive interaction, $\gamma =3$. As in the case of
the HO potential, one can clearly see the GS phase transition caused by the
energy-level inversion, the transition points being $\beta _{0}=0.78$, $%
\beta _{1}=1.77$, $\beta _{2}=3.08$, \textit{etc}. At $\beta >\beta _{1}$,
the GS again develops a pattern in the form of a spatially confined lattice,
in both components $\psi _{1}$ and $\psi _{2}$. The size of the lattice is $%
L=3$, which coincides with the width of the box potential, and the period
can be approximated by $T_{n}=2L/n$.

For $\gamma =0$, the distribution of atoms reveals new results in Figs.~\ref%
{figure8}(c,d). Unlike the superposition state generated in the case of the
HO potential, the atoms are distributed at both top and bottom edges of the
box. With the increase of strength $\gamma $ of the inter-component
attraction, the symmetry defined by Eq.~\eqref{symmetriy} gets broken, and
the GS takes the form of an edge state. The top edge states for $\gamma =-1.5
$ and $\gamma =-4$ are shown in Figs.~\ref{figure8}(e,f) and (g,h),
respectively. One can see that there is a region of transition of the
eigenstate towards the edge state, in the range of $0<\beta <0.5$ for $%
\gamma =-1.5$, while the transition region almost disappears at $\gamma =-4$%
. The bottom edge states (not shown here) can be obtained from their
top-edge counterparts, by substitution~\eqref{substitution}.

The results indicate that the energy-level inversion and edge states are
chiefly generated by the combined effect of the gradient magnetic field and
SO coupling, while the particular shape of the trapping potential affects
profiles of the GS\ wave function and the position of the phase-transition
point, $\beta _{n}$.
\vspace{3em}

\section{Conclusion}

We have proposed a method to realize the tunable energy-level inversion in
the SO-coupled BEC. The binary condensate is trapped in the HO
(harmonic-oscillator) potential, and is subject to the action of the
gradient magnetic field, which results in the energy-level quantization. By
introducing the shifted quantum-number density operator, the linear version
of the system can be solved exactly, in terms of the Hermite-Gaussian
functions. By adjusting the SO coupling strength and magnetic-field
gradient, the inversion of the energy levels occurs, making it possible to
transform any bound state into the GS. Stationary solutions of the full
nonlinear system are obtained numerically. In the case of the
inter-component repulsion, the numerical results follow the pattern of the
linear eigenfunctions. In the case of the inter-component attraction, the GS
takes the form of superposition and edge states, at different values of the
SO-coupling strength. Replacing the HO trap by the box potential, we have
checked that the energy-level inversion and edge states are chiefly
generated by the combined effect of the gradient magnetic field and SO
coupling.

\section*{ACKNOWLEDGMENTS}

This research was supported by the 111 project (grant No. D18001), the
Hundred Talent Program of the Shanxi Province (2018), the National Key R\&D
Program of China (grants No. 2021YFA1400900, 2021YFA0718300,
2021YFA1400243), NSFC (Nos. 61835013, 12234012), Space Application System of China 
Manned Space Program, and by the Israel Science foundation (grant No. 1695/22).

\end{document}